\documentclass[showpacs, 12pt]{revtex4}

\usepackage{graphicx}%
\usepackage{dcolumn}
\usepackage{amsmath}

\begin{document}

%\preprint{HEP/123-qed}

\title{\bf Direct evidence for the validity of Hurst's empirical law in hadron production
processes}
%\footnote{Supported
%by the National Natural Science Foundation of China under Grant No.70271064.}}% Force line breaks with \\

% \email{liuq@ccnu.edu.cn}
%\author{ MENG Ta-chung$^{1,2}$\footnote{Email address: meng@ccnu.edu.cn; meng@physik.fu-berlin.de}}%
 % \email{meng@ccnu.edu.cn  meng@physik.fu-berlin.de}
%\author{LIU Qin$^{1}$\footnote{Email address: liuq@ccnu.edu.cn}}
%\affiliation{\it $^{1}$Department of Physics, CCNU, 430079 Wuhan,
%China
%\\ $^{2}$Institut f\"{u}r Theoretische Physik, FU-Berlin, 14195 Berlin, Germany}

\author{LIU Qin \footnote{Email address: liuq@mail.ccnu.edu.cn}}
\affiliation{Department of Physics, CCNU, 430079 Wuhan, China}
\author{MENG Ta-chung \footnote{Email address: meng@mail.ccnu.edu.cn; meng@physik.fu-berlin.de}}
 %%\homepage{http://www.Second.institution.edu/~Charlie.Author}
\affiliation{Department of Physics, CCNU, 430079 Wuhan, China}
\affiliation{Institut f\"{u}r Theoretische Physik, FU-Berlin,
14195 Berlin, Germany}
%%Second institution and/or address\\

%%This line break forced% with \\
%%}%

\date{\today}% It is always \today, today, but you may specify any date with \date.

\begin{abstract}
We propose to use the rescaled range analysis to examine the
records of rapidity-dependence of multiplicities in high-energy
collision processes. We probe event by event the existence of
global statistical dependence in the system of produced hadrons,
and demonstrate the effectiveness of the above-mentioned
statistical method by applying it to the cosmic-ray data of the
JACEE collaboration, and by comparing the obtained results with
other experimental results for similar reactions at accelerator
and collider energies. We present experimental evidence for the
validity of Hurst's empirical law, and the evidence for the
existence of global statistical dependence, fractal dimension, and
scaling behavior in such systems of hadronic matter. None of these
features is directly related to the basis of the conventional
physical picture. Hence, it is not clear whether (and if yes, how
and why) these striking empirical regularities can be understood
in terms of the conventional theory.
\end{abstract}

% PACS, the Physics and Astronomy Classification Scheme.

\maketitle

%%\tableofcontents
It is known since decades that hadrons can be produced through
energy conversion in various kinds of high-energy collisions. Yet,
not much is known about the mechanism(s) of such production
processes. The conventional way of describing/understanding the
formation process of such hadrons is as follows \cite{1,2}. One
starts with the generally accepted basic constituents (quarks,
antiquarks and gluons) of matter and the set of rules (QCD
Lagrangian, Color Confinement, running coupling constants etc.)
which describe how such entities interact with one another. This
is often considered to be ``the first stage'' of describing the
formation process in the conventional picture, although in
practice, these basic constituents inside the colliding objects
are often considered to be approximately free. The discussion on
the dynamics of hadron-production begins actually when the
interactions/correlations are taken into account between the basic
constituents at the quark-gluon-level and/or those between hadrons
which are considered to be ``the basic constituents'' at the next
level. At this, ``the second stage'', some people prefer to
examine the formation and disintegration of resonances and/or
clusters which are made out of hadrons. Such objects are either
taken to be experimentally known short-lived hadrons or
calculated/parametrized by using theoretical models. Correlations
between the produced hadrons are often described by correlation
functions in the same manner as those in the cluster expansion
technique of Ursell and Mayer \cite{3}. In order to include the
effects of higher order correlations and/or multiparticle
interactions in general and concentration of hadrons in small
kinematical regions in ``spike events'' in particular, many people
use the method of factorial moments suggested by Bialas and
Peschanski \cite{4} based on a two-component picture \cite{5} for
hadron-production. Alternatively, other people prefer to study,
immediately after the first stage, the formation of ``quark-gluon
plasma'' and the mechanism(s) which turn such plasma into
measurable hadrons \cite{2}.

The basic difficulties encountered by the conventional concepts
and methods described above are two-fold. (a) Quarks, antiquarks
and gluons have not been, and according to QCD and Color Confiment
they can never be, directly measured. (b) Pertubative methods for
QCD-calculations (pQCD) can be used only when the momentum
transfer in the scattering process is so large that the
corresponding QCD running coupling constant is less than unity,
but the overwhelming majority of such hadron-production processes
are ``soft'' in the sense that the momentum transfer in such
collisions is relatively low. This implies that comparison between
the calculated results and the experimental data for the
measurable hadrons can be made only when various assumptions and a
considerable number of adjustable parameters are introduced
\cite{2}.

Having these facts in mind, we are naturally led to the following
questions. Do we really {\it need} all the detailed information
mentioned above, which contains so many assumptions and adjustable
parameters, to find out what {\it the key features} of high-energy
hadron production processes are? For the purpose of
describing/understanding such a process, is it possible to take a
global view of hadron production by looking at it simply as a
process of energy conversion into matter, by dealing only with
quantities which can be directly measured, and by working only
with assumptions which can be checked experimentally? Attempts to
understand the mechanism(s) of high-energy hadron production
through data-analyses by using statistical methods have been made
already in the 1970's. As a typical example we discuss the work by
Ludlam and Slansky \cite{6,7} (thereafter referred to as the
LS-approach), and other related papers cited therein. The common
goal of the LS-approach and our approach is: First of all, clearly
the ultimate goal of performing such data-analyses. Namely to
extract, as directly as possible, useful information about the
general features of the reaction mechanism(s) of multihadron
production processes in high-energy hadronic reactions. Second,
common to both approaches is also the examination of fluctuation
phenomena in an event-by-event manner, especially those in the
longitudinal variables of such production processes. There are,
however, also vast differences between the two approaches: While
the main purpose of the LS-approach is to study {\it clustering
effects}, where particular emphasis is given to the estimation of
the size of the emitted clusters in exclusive or semi-inclusive
reactions (in order to avoid the effect of kinematical
constrains). Such clusters are assumed to be produced through
independent emission. This means, while ``relatively short range
correlations'' between the observed hadrons are taken into account
through the existence of hadronic clusters, the question whether
global statistical dependence (also known as ``long-run
statistical dependence'') exists in the longitudinal variables has
been left open. Complimentary to the LS-approach, the main concern
of {\it our} approach is {\it to probe the existence of such
global statistical dependence} by using experimental and only
experimental data. To be more precise, the purpose of this series
of research (see also Ref. \cite{8}) is try to extract useful
information on the reaction mechanism(s) of such processes by
using a {\it preconception-free} data-analysis, namely, (i) {\it
without} assuming that we know all the dynamical details about the
basic constituents and their interactions, (ii) {\it without}
applying pertuabtive methods to QCD or using phenomenological
models for doing calculations, and (iii) {\it without} assuming
that {\it only} statistical methods which lead to {\it finite
variances} and {\it local} (short-run, e.g. Markovian) statistical
dependence are valid methods.

We recall that, the two assumptions mentioned in (iii), namely
{\it finite variances} and {\it local statistical dependence},
have always been {\it a matter of course} in practical statistics.
But, as it is known since the 1960's that a large amount of
heavy-tailed empirical records have been observed in various
fields and they can be best interpreted by accepting {\it
infinite} variances \cite{9,10,11}. While most familiar examples
are found in finance and economics \cite{9,10}, striking examples
have also been observed in hadron-production processes: the
relative variations of hadron-numbers between successive rapidity
intervals are shown \cite{8} to be {\it non-Gaussian} stable
random variables which exhibit stationarity and scaling. Taken
together with the fact that the main statistical technique to
treat very global statistical dependence is spectral analysis
which performs poorly on records which are far from being Gaussian
\cite{9,10,11}, we propose to use a more general statistical
method to examine the rapidity-distributions of the produced
hadrons in high-energy collisions: the rescaled range analysis.
This method was originally invented by Hurst \cite{12}, a
geophysicist, who wanted to design an ideal reservoir which never
overflows and never empties; and was later mathematically
formalized and developed by Mandelbrot and his collaborators
\cite{10,11,13} into an extremely powerful statistical method.

What Hurst had was the record of observed annual discharge,
$\xi(t)$, of Lake Albert for the total period of 53 years, where
$t$ is a discrete integer-valued time between some fixed starting
point $t_{0}$ and some time-span $\tau$ within the total time
period considered. The time-span $\tau$ is known as the {\it lag}
in the literature \cite{12,13,14}. By requiring that the reservoir
should release a regulated volume each year which equals to the
average influx
\begin{equation}
\langle\xi\rangle_{t_{0}, \tau}=\frac{1}{\tau}\sum
\limits_{t=t_{0}}^{t_{0}+\tau}\xi(t),
\end{equation}
the accumulated departure of the influx $\xi(t)$ from the mean
$\langle\xi\rangle_{t_{0},\tau}$ is
\begin{equation}
X(t_{0},t,\tau)=\sum
\limits_{u=t_{0}}^{t_{0}+t}\{\xi(u)-\langle\xi\rangle_{t_{0},\tau}\}.
\end{equation}
The difference between the maximum and the minimum accumulated
influx is
\begin{equation}
R(t_{0},\tau)=\max_{0\leq t\leq \tau} X(t_{0},t,\tau)-\min_{0\leq
t\leq \tau} X(t_{0},t,\tau).
\end{equation}
Here, $R(t_{0},\tau)$ is called the {\it range}, and it is nothing
else but the storage capacity required to maintain the mean
discharge throughout the lag $\tau$.  It is clear that the range
depends on the selected starting point $t_{0}$ and the lag $\tau$
under consideration. Noticing that $R$ increases with increasing
$\tau$, Hurst examined in detailed the $\tau$-dependence of $R$.
In fact, he investigated not only the influx of a lake but also
many other natural phenomena. In order to compare the observed
ranges of these phenomena, he used a dimensionless ratio which is
called the {\it rescaled range}, $R/S$, where $S(t_{0},\tau)$
stands for the sample standard deviation of record $\xi(t)$
\begin{equation}
S(t_{0},\tau)=\left ( \frac{1}{\tau}\sum
\limits_{t=t_{0}}^{t_{0}+\tau}
\{\xi(t)-\langle\xi\rangle_{t_{0},\tau}\}^{2}\right )^{1/2}.
\end{equation}
The result of the comparison is that the $\tau$-dependence of the
observed rescaled range, $R/S$, for many records in nature is well
described by
\begin{equation}
\frac{R}{S}(t_{0},\tau)=(\frac{\tau}{2})^{H}.
\end{equation}
This simple relation is now known as {\it Hurst's Empirical Law}
\cite{10,11,12,13,14}, and $H$, the {\it Hurst exponent}
\cite{10,11,12,13,14}, is a real number between 0 and 1. This
powerful method of testing the relationship between the rescaled
range $R/S$ and the lag $\tau$ for some fixed starting point
$t_{0}$ is called the rescaled range analysis, also known as the
$R/S$ analysis \cite{10,11,12,13,14}.

The rescaled range, $R/S(t_{0},\tau)$, is a very robust statistic
for testing the presence of global statistical dependence. This
robustness extends in particular to processes which are
extraordinarily far from being Gaussian. Furthermore, the
dependence on $\tau$ of the average of the sample values of
$R/S(t_{0},\tau)$, carried over all admissible starting points,
$t_{0}$, within the sample, $\langle
R/S(t_{0},\tau)\rangle_{t_{0}}\sim \tau^{H}$, can be used to test
and estimate the {\it $R/S$ intensity} $J=H-1/2$. The special
value $H=1/2$, and thus $J=0$, corresponds to {\it the absence} of
global statistical dependence, and it is characteristic of
independent, Markov or other local dependent random processes.
Positive intensity expresses persistence, negative intensity
expresses antipersistence. Another important aspect of this
statistical method is its {\it universality}. As can be readily
seen, the $R/S$ analysis is {\it not only} useful for the design
of ideal reservoirs; and it is {\it not only} applicable when time
records are on hand. In fact, the ideal reservoir is nothing else
but a device of quantifying the measurements of some phenomena in
Nature, where time simply plays the role of an ordering number.

Can the recorded rapidity-distributions of produced hadrons, for
example the quantity discussed in Ref. \cite{8}, be used as
ordered records for R/S analysis where the rapidity plays the role
of an ordering number? While the total kinematically allowed
rapidity interval ($Y^{tot}_{max}$) in a single-event is uniquely
determined by the total c.m.s energy $E_{cms}\equiv \sqrt{s}$ of
the corresponding collision process as well as the masses of the
colliding objects, and the collision processes under consideration
are approximately symmetric with respect to c.m.s. because the
differences between the masses are negligible compared to the
kinetic energies, we denote by $y$ the rapidity of an observed
hadron, and consider, as we did in Ref. \cite{8}, the
rapidity-dependent quantity, $\ln dN/dy(y)$, as the ordered record
within a chosen symmetric rapidity interval of a colliding system,
$Y_{max}\equiv y_{f}-y_{i}$, which is the rapidity interval
measured from some initial value $y_{i}$ to a final value $y_{f}$
where $|y_{i}|=y_{f}=Y_{max}/2$. It is clear that $Y_{max}\leq
Y^{tot}_{max}$ defined above, and that the lag $Y$ which is the
counterpart of the period $\tau$ in the case of Lake Albert varies
between zero and $Y_{max}$. For a given $Y$ belonging to the
$Y_{max}$, the total averaged multiplicity of the collision
process within the lag $Y$ is
\begin{equation}
\langle \ln\frac{dN}{dy} \rangle _{y_{i},Y}=\frac{1}{Y} \sum
\limits_{y=y_{i}}^{y_{i}+Y} \ln \frac{dN}{dy}(y).
\end{equation}
The accumulated departure of $\ln dN/dy(y)$ from the mean $\langle
\ln dN/dy \rangle_{y_{i},Y}$ is
\begin{equation}
X(y_{i},y,Y)=\sum \limits_{u=y_{i}}^{y_{i}+y}
\{\ln\frac{dN}{dy}(u)-\langle \ln\frac{dN}{dy}
\rangle_{y_{i},Y}\},
\end{equation}
and the corresponding range $R(y_{i},Y)$ is
\begin{equation}
R(y_{i},Y)=\max_{0 \leq y \leq Y}X(y_{i},y,Y)-\min_{0 \leq y \leq
Y}X(y_{i},y,Y).
\end{equation}
Here, $R(y_{i},Y)$ represents the difference between the maximum
and the minimum deviation of the amount of energy in form of
number of hadrons with average energy $\epsilon$ which can never
be larger than the total (c.m.s) energy $\sqrt{s}$, nor be less
than zero. Dividing $R(y_{i},Y)$ by the corresponding sample
standard deviation
\begin{equation}
S(y_{i},Y)=\left ( \frac{1}{Y} \sum
\limits_{y=y_{i}}^{y_{i}+Y}\{\ln\frac{dN}{dy}(y)-\langle
\ln\frac{dN}{dy} \rangle_{y_{i},Y}\}^{2}\right )^{1/2},
\end{equation}
we thus obtain the rescaled range $R/S(y_{i},Y)$ for the given $Y$
in the high-energy hadron production process.

Having seen the motivations of performing such kind of analysis,
and the fact that there exists no technical problems in carrying
out it, we repeat the above-mentioned procedures for all
admissible $Y$'s within the chosen rapidity interval $Y_{max}$. We
are now ready to check whether or not it is true that
\begin{equation}
\frac{R}{S}(y_{i},Y)\sim Y^{H(y_{i})},
\end{equation}
where the corresponding Hurst exponent is indeed a real number
between zero and unity.

As illustrative examples, we consider the two well-known
cosmic-ray events measured by JACEE-collaboration \cite{15}. We
recall that the JACEE-data have attracted much attention
\cite{2,4,5,8} not only because they are taken at energies much
higher than those taken at accelerator/collider energies (and thus
are usually associated with high-multiplicity events), but also
because they exhibit significant fluctuations. Here we use, as in
Ref. \cite{8}, the short-hand JACEE1 and JACEE2 for the Si+AgBr
collision at 4 Tev/nucleon and the Ca+C (or O) collision at 100
Tev/nucleon respectively.

First, we probe the dependence of $R/S(y_{i},Y)$ on the lag $Y$
for some fixed initial value $y_{i}$. Having in mind that
pseudorapidity $\eta$ is a good approximation of rapidity $y$, we
calculate the quantity, $R/S(y_{i},Y)$, on the left-hand-side of
Eq. (10) for the following rapidity intervals. In JACEE1, we take
$y_{i}=-4.0$ and thus $Y_{max}=80$, correspondingly we take
$y_{i}=-5.0$ in JACEE2 hence $Y_{max}=100$, where in both JACEE
events we have taken the experimental resolution power, 0.1, into
account, and $Y$ is measured in units of this resolution power.
The obtained results of this check are shown in Figs. (1a) and
(1c) respectively. In these $\log-\log$ plots, the slope of the
data points (shown as black dots for JACEE1 and black triangles
for JACEE2) determines the Hurst exponent $H$. It is approximately
0.9 in both cases (indicated by the solid lines). For the sake of
comparison, we plot in the same figure two samples (the
sample-size is taken to be the same as that of JACEE1 and JACEE2
respectively) of {\it independent} Gaussian random variables. As
expected, the corresponding points which are shown as open circles
and triangles respectively lay on straight lines (indicated by
broken lines) with slopes approximately equal to 0.5. In this
connection, it is of considerable importance to mention that the
following especially the relationship between ``$H=0.5$'' and
``Gaussian distribution'' has been discussed in detail by several
authors in particular by Mandelbrot (see e.g. p. 387 of Ref.
\cite{11} and the papers cited therein. Note that in Ref.
\cite{11} Mandelbrot uses $J$ for the Hurst exponent and used $H$
for the exponent associated with fractional Brownian motion). It
has been shown that $H=0.5$ is valid for {\it independent} random
processes with or without finite variance. A typical example for
the former case (finite variance) is (independent) Gaussian, and
the most well-known example for the latter (infinite variance) is
the white L\'{e}vy stable noise. In other words, since there are
independent and dependent Gaussian random processes (an example
for the latter can, e.g., be the distributions associated with
fractional Brownian motion \cite{9,10,11}) the Hurst exponent for
Gaussian may or may not be 0.5. It should also be mentioned that
the preconception-free method for data-analyses discussed in a
previous paper \cite{8} is indeed able to test and uniquely
determine whether the distribution under consideration is
Gaussian. But the problem whether it is independent or dependent
remains unresolved and will be answered in this paper.

In order to compare the present as well as the previously
suggested \cite{8} method with experiments performed for similar
collision processes at accelerator and collider energies, we did
the following. Although it was very hard to find, and it indeed
took a very long time to find them, we nevertheless prefer to use
{\it real data}. This is because we think {\it real data} for
similar collision processes should be more reliable than Monte
Carlos simulations for various processes, namely, we cannot know
for sure what kind of preconceptions have been built-in the codes
of such simulations. No published data could be found. But
fortunately enough, some former EMU01 group members agreed to
analyze their yet unpublished high-energy high multiplicity and
large fluctuation data, and also analyzed such kind of data given
to them by the STAR-Collaboration at RHIC measured in the limited
pesudorapidity range $-1\leq\eta\leq 1$. We are very grateful to
LAN {\it et al.} \cite{16} who generously show some of their test
results before the publication, and allow us to quote part of
them. We also wish to thank EMU01 and STAR Collaboration for
allowing LAN {\it et al.} to use their data.

LAN {\it et al.} \cite{16} analyzed about 20 EMU01 events taken at
CERN at lab-energy 200 AGeV in S$^{32}$+Au$^{197}$ reactions. The
rapidity range is from about $\eta_{lab}=-1$ to about
$\eta_{lab}=7.0$ and the bin-size $\Delta\eta$ is taken to be 0.2
in order to avoid empty bins. The lowest and the highest
multiplicities of charged hadrons are 199 and 260 respectively.
Their result shows that Hurst's law is satisfied in 99.5$\%$ of
the analyzed events. The observed values for $H$ are approximately
the same as those obtained from JACEE-events, and that all of the
observed $H$ values are definitely much larger than 0.5. See, for
example, Figs. (2a) and (2c). It should of some interest to point
out that LAN {\it et al.} \cite{16} also applied the method
proposed in Ref. \cite{8} to check whether the events under
consideration are Gaussian. It is found that 61$\%$ of them are
{\it non-Gaussian}.

LAN {\it et al.} \cite{16} also analyzed the relativistic
($\sqrt{s_{NN}}=200$ GeV) heavy-ion data obtained from STAR
Collaboration. See Figs. (3a) and (3c). The $\eta$-range is rather
small ($-1\leq\eta\leq 1$), but since the energy is high, the
multiplicities of charged hadrons are nevertheless high (up to
959) and many of the events do exhibit large fluctuations. It is
seen \cite{16} that among the 50 analyzed events, 82$\%$ of them
show that Hurst's law is valid where all the obtained Hurst
exponents are approximately 0.6. Furthermore, it is also seen
\cite{16} that 84$\%$ of the analyzed $\eta$-distributions are
{\it non-Gaussian}.

Second, we change the initial values $y_{i}$ and test whether/how
the validity of Hurst's empirical law and whether/how the
corresponding values of the Hurst exponent $H(y_{i})$ depend on
$y_{i}$ in Eq. (10). We consider in JACEE1: $y_{i}=-0.5$, -1.0,
-2.0 and -3.0, where the corresponding values of $y_{f}$ are such
that $Y_{max}=1.0$, 2.0, 4.0 and 6.0 respectively; and in JACEE2:
$y_{i}=-0.5$, -1.0, -2.0, -3.0 and -4.0 such that $Y_{max}=1.0$,
2.0, 4.0, 6.0 and 8.0 respectively. Two of these plots, namely for
$y_{i}=-0.5$ thus $Y_{max}=1.0$ for both JACEE events, are shown
in the Figs. (1b) and (1d). What we see is that $H$ is {\it
independent} of $y_{i}$ because the values of $H$ remain
approximately the same as those for the corresponding $H$ values
without any change of $y_{i}$. The possible existence of effects
caused by changing the initial values have not only been probed
for JACEE events but also for the above-mentioned EMU01 and STAR
data (see Ref. \cite{16} and the papers cited therein). The
results of which are also shown in the Figs. (2b), (2d) and (3b),
(3d). This means, also the independence on $y_{i}$ remains at
lower energies.

Several conclusions can be drawn directly from the figures.

First, the obtained results show evidence for the validity of
Hurst's empirical law in high-energy hadron production processes.
The values of the Hurst exponent $H$ extracted from the two JACEE
events \cite{15} are approximately {\it the same}, although
neither the total c.m.s energies nor the projectile-target
combinations in these colliding systems are the same. The results
of LAN {\it et al.} \cite{16} show that such characteristics
retain also at accelerator and collider energies provided that the
multiplicities are not too small although the size of the limited
rapidity range (e.g. $|\eta|\leq 1$ for STAR) has some influence
on the absolute values of $H$, namely 0.6 instead of 0.9. This
seems to suggest once again \cite{8} the possible existence of
universal features when the data are analyzed in a
preconception-free manner.

Second, the fact that not only the scaling behavior of $R/S$, but
also the value of $H$ is independent of $y_{i}$ shows once again
\cite{8} that the process is stationary.

Third, the Hurst exponent found here is greater than 1/2 which
implies the existence of {\it global statistical dependence} in
the system of produced hadrons in such collision processes and
thus the existence of {\it global structure} discussed in detail
by Mandelbrot and his collaborators \cite{9,12}.

Fourth, having in mind that rapidity is defined with respect to
the collision axis, and that the momenta of the produced hadrons
along this direction are very much different with those in the
perpendicular plane, the fractal structure, if it exists, is
expected to have its geometric support along this axis and is
self-affine.

Last but not least, the obtained Hurst exponent $H$ can be used to
determine the fractal dimensions $D_{G}=2-H$ and $D_{T}=1/H$
defined by Mandelbrot (see Ref. \cite{9}, p. 37). For both JACEE
events \cite{15} we have $D_{T}\approx D_{G}=1.1$.

Furthermore, it should be pointed out that the validity of the
universal power law behavior of the extremely robust quantity
$R/S$ shown by the independence of $H$ on $y_{i}$ (which implies
independence of $H$ on $Y_{max}$) in Eq. (10) and in the figures
strongly suggests the following. There is no intrinsic scale in
the system in which the hadrons are formed. Taken together with
the uncertainty relation, $\Delta y\Delta l\sim$ constant,
discussed in detail in Ref. \cite{8} where $\Delta l$ stands for
locality, the canonical conjugate of $y$ in light-cone variables
in space-time \cite{17}, the statement made above is true also
when the hadron-formation process is discussed in space-time.

In conclusion, the powerful statistical method, $R/S$ analysis,
has been used to analyze the JACEE-data \cite{15}. Taken together
with the results of LAN {\it et al.} \cite{16} we are led to the
conclusion that direct experimental evidence for the existence of
global statistical dependence, fractal dimension, and scaling
behavior have been obtained. Since none of these features is
directly related to (the entirety or any part of) the basis of the
conventional picture, it is not clear whether, and if yes, how and
why these striking empirical regularities can be understood in
terms of the conventional theory.

%\begin{acknowledgments}
The authors thank LAN Xun and LIU Lei for helpful discussions and
KeYanChu of CCNU for financial support. This work was supported by
the National Natural Science Foundation of China under Grant No.
70271064 and 90403009.
%\end{acknowledgments}

\begin{figure}
\includegraphics[width=0.7\textwidth]{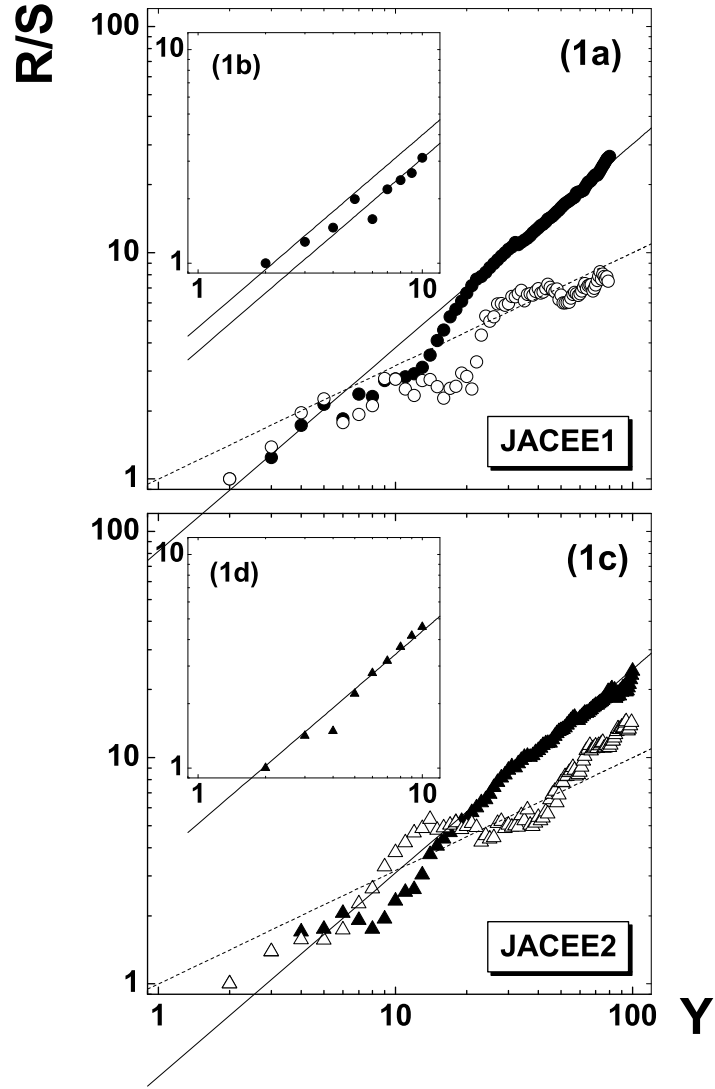}
\caption{$R/S(y_{i},Y)$ as functions of $Y$ for given $y_{i}$'s;
data are taken from Ref. \cite{15}. For details see text.}
\label{fig1}
\end{figure}

\begin{figure}
\includegraphics[width=0.7\textwidth]{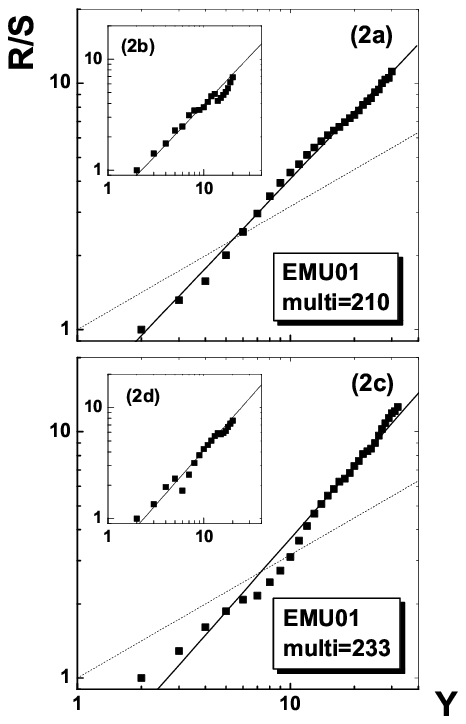}
\caption{$R/S(y_{i},Y)$ as functions of $Y$ for given $y_{i}$'s.
The results are taken from LAN {\it et al.} \cite{16}. The broken
lines indicate the $H=0.5$ case.} \label{fig2}
\end{figure}

\begin{figure}
\includegraphics[width=0.7\textwidth]{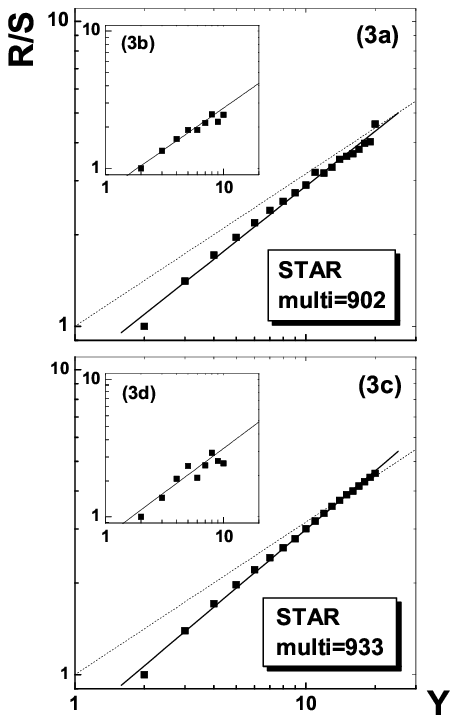}
\caption{$R/S(y_{i},Y)$ as functions of $Y$ for given $y_{i}$'s.
The results are taken from LAN {\it et al.} \cite{16}. The broken
lines indicate the $H=0.5$ case.} \label{fig3}
\end{figure}

%\newpage %J

\bibliography{apssamp}% Produces the bibliography via BibTeX.

%\newpage
%\begin{figure}[tbph]
%\includegraphics[width=0.45\textwidth]{fig1}
%\caption{}
%\label{fig1}
%\end{figure}

\end{document}